\documentclass[a4paper,runningheads]{llncs}
\usepackage[T1]{fontenc}
\usepackage{lmodern}
\usepackage{graphicx}
\usepackage{amsmath,amssymb}
\usepackage{booktabs}
\usepackage{array}
\usepackage{tikz}
\usetikzlibrary{positioning,arrows.meta,calc,shapes.geometric,fit}
\usepackage[protrusion=true,expansion=false]{microtype}
\usepackage{algorithm}
\usepackage[noend]{algpseudocode}
\usepackage{enumitem}
\usepackage[hidelinks]{hyperref}
\hypersetup{pdftitle={Ontology-Governed Epistemic Authorization in Heterogeneous LLM Multi-Agent Systems},pdfauthor={Tao Shan, Huifeng Zhang, Jian Liu}}
\begin{document}
\title{Ontology-Governed Epistemic Authorization in Heterogeneous LLM Multi-Agent Systems}
\titlerunning{Ontology-Governed Epistemic Authorization}
\author{Tao Shan \and Huifeng Zhang \and Jian Liu}
\authorrunning{T. Shan et al.}
\institute{University of Birmingham, Birmingham, United Kingdom\\
\email{txs536@student.bham.ac.uk}, \email{hxz451@student.bham.ac.uk},\\
\email{j.liu.22@bham.ac.uk}}
\maketitle

\begin{abstract}
When several large language model (LLM) agents read the same evidence, they often disagree, and a multi-agent system (MAS) must merge their proposals into one shared state before anyone acts on it. The merge step is usually an implicit vote or an overwrite, and it can turn an unresolved question into a confident answer. We present a protocol that makes this step explicit and governable. Agents emit typed proposals in which \textsc{Unknown} and \textsc{Conflicting} are first-class values; a single credentialed role writes the shared state through a named transformation; and a completion-aware gate authorizes an action only when every way of filling the unresolved fields leads the policy to the same action, abstaining otherwise. We prove that a transformation which keeps more disagreement can never authorize more actions, and we give the conditions under which the gate is sound. Three controlled evaluations separate the contributions of the parts. Executing an explicit policy over the typed state, rather than asking an LLM to apply it, raises exact action recovery from 24.2\% to 100\%; the schema-declared authority relation blocks 72/72 cross-role writes; and on a fixed replay of 36 recorded proposal profiles for Common Vulnerabilities and Exposures (CVE) triage, preserving disagreement gives 0/12 false authorizations at 0.708 coverage, against 1/12 at 0.833 for field-wise majority vote and 2--4/12 for last-writer-wins, depending only on arrival order. The evidence is limited to one replay profile and a policy-derived oracle; we state these limits and the engineering lessons that follow from them.

\end{abstract}

\section{Introduction}\label{sec:intro}

\paragraph{The problem.} LLM-based multi-agent systems coordinate model instances through agent profiles and communication protocols~\cite{guo2024llmmultiagents}. A common pattern assigns several agents to read the same evidence, merges their proposals into one shared state, and lets a downstream role act on that state. Running several agents is meant to make the result more trustworthy, but it does not do so by itself. When agents produce similar responses, multi-LLM debate can converge on a shared majority opinion, including a shared misconception~\cite{estornell2024multillm}. When the agents disagree, the disagreement is information, and the step that merges their proposals decides whether that information survives until someone acts.

\paragraph{A running example.} Consider vulnerability triage, the setting of our main evaluation. Three agents, backed by three different LLMs, read the same advisory and fill a five-field record; one field is \emph{exploit maturity} (EM). On one case of our benchmark, two agents report EM${}={}$\textsc{Confirmed} and the third reports \textsc{Unknown}. A field-wise majority vote writes \textsc{Confirmed}, the record now looks complete, and the triage policy returns \textsc{ImmediatePatch}. The benchmark, however, labels this case as one where the system should abstain, because the evidence on EM is in conflict and the policy assigns different actions to different EM values. The same three proposals, merged by a rule that writes \textsc{Conflicting} whenever the agents do not agree, lead the gate to see that the unresolved record is consistent with three different actions, so it abstains. Nothing differs between the two runs except the merge step. Section~\ref{sec:example} follows this case through every component of the protocol.

\paragraph{Why existing machinery does not settle it.} Each part of this problem has been studied, but in a different place in the pipeline. Answer-level aggregation such as voting, score fusion, or debate~\cite{du2024multiagentdebate} resolves proposals into a single output; it does not keep the disagreement for a later decision. Normative institutions and role-based access control~\cite{esteva2001electronic,dignum2004model,sandhu1996rbac} govern \emph{who} may write the shared state, not whether the written state warrants action. Selective prediction and LLM abstention~\cite{geifman2019selectivenet,wen2025abstention} decide when a model should decline to answer, usually from the model's own confidence, not from the gap between a shared state and an explicit decision policy. Runtime verification can return an inconclusive verdict over a finite trace~\cite{bauer2011runtime}, but it monitors temporal properties and does not prescribe how the monitored state is built from conflicting proposals. In our example, none of these mechanisms on its own distinguishes the two runs: both have an authorized writer, both produce a confident-looking record, and the LLMs involved are identical.

\paragraph{Our approach.} We treat the merge step, which we call the \emph{organizational transformation}, as a named and governed part of the MAS. The protocol has three parts. (i)~Agents emit \emph{typed} proposals in which \textsc{Unknown} and \textsc{Conflicting} are values of each field, distinct from each other and from concrete values. (ii)~A single credentialed role applies an explicit transformation $T$ and is the only role allowed to write the shared evidence state; every write is checked against a schema-declared role--graph relation and logged. (iii)~A deterministic gate enumerates all \emph{completions} of the written state, i.e.\ all ways of replacing its unresolved fields by concrete values, applies the policy to each, and authorizes an action only if all completions lead to the same one. The shared state lives in a Resource Description Framework (RDF) store with Web Ontology Language (OWL)~2 vocabularies and Shapes Constraint Language (SHACL) constraints, so that vocabularies, authority, and provenance are declared rather than hard-coded; a typed dictionary can implement the same decision semantics (Section~\ref{sec:spec}).

\paragraph{What is new.} We do not claim that typed values, access control, or abstention are new in isolation. What the composition adds is a single ordering in which merge rules can be compared \emph{before} they are run: if one transformation always leaves a superset of completions compared with another, it can never authorize an action the other would not (Theorem~A). This turns the choice of merge rule into an analysable design decision with a predictable safety--coverage consequence, and the protocol records enough of each run to check that prediction case by case. Our contributions are:
\begin{itemize}[leftmargin=*,itemsep=1pt,topsep=2pt]
\item a typed, single-writer LLM-MAS protocol with a completion-aware authorization gate, and a per-case procedure (Algorithm~\ref{alg:loop}) that an engineer can reimplement without the RDF substrate;
\item a completion-refinement ordering over organizational transformations with three conditional guarantees: authorization antitonicity, invariance under rejected writes, and order invariance of a join-based comparator that is nonetheless unsafe;
\item a controlled evaluation in which proposals, policy, gate, and case set are held fixed while only one component varies, separating the effects of vocabulary, executable policy, authority enforcement, and transformation;
\item engineering lessons for MAS designers and a list of what the present evidence does not establish (Sections~\ref{sec:lessons} and~\ref{sec:threats}).
\end{itemize}

Three research questions (RQ1--RQ3), examined in evaluations E1--E3, organize the analysis.

\textbf{RQ1/E1.} Does an explicit institutional vocabulary improve state coordination, and what is the contribution of policy availability?

\textbf{RQ2/E2.} Does completion-aware authorization distinguish actionable from unresolved states, independently of RDF serialization?

\textbf{RQ3/E3.} When heterogeneous proposals are transformed organizationally, does preserving disagreement avoid unsafe action, and what are the order and common-mode boundaries?

Section~\ref{sec:example} walks through the running example and the per-case procedure; Section~\ref{sec:protocol} formalizes the protocol and its guarantees; Section~\ref{sec:related} positions the work; Sections~\ref{sec:eval-design}--\ref{sec:e3} report the evaluations; Section~\ref{sec:discussion} discusses the results, engineering lessons, threats to validity, and future work.

\section{Overview and Running Example}\label{sec:example}

\begin{figure}[t]
\centering
%
\begin{tikzpicture}[
    font=\scriptsize,
    agent/.style={rectangle, draw, rounded corners=1pt,
                  minimum width=1.75cm, minimum height=0.62cm,
                  align=center, fill=gray!8, inner sep=1.5pt},
    pool/.style={rectangle, draw, rounded corners=1pt,
                 minimum width=1.05cm, minimum height=1.05cm,
                 align=center, fill=gray!15},
    tbox/.style={rectangle, draw, rounded corners=1pt,
                 minimum width=0.98cm, minimum height=0.44cm,
                 align=center, fill=white, inner sep=1pt},
    gate/.style={diamond, draw, thick, aspect=1.85, inner sep=0pt,
                 align=center, fill=white},
    outact/.style={rectangle, draw, rounded corners=1pt,
                   minimum width=1.30cm, minimum height=0.44cm,
                   align=center, fill=white, inner sep=1pt},
    outabs/.style={rectangle, draw, rounded corners=1pt,
                   minimum width=1.30cm, minimum height=0.44cm,
                   align=center, fill=gray!28, inner sep=1pt},
    arr/.style={-{Stealth[length=1.6mm]}, semithick},
]

\node[agent] (q) at (-5.15,  0.78) {\textsc{ag\_q}\\Qwen};
\node[agent] (d) at (-5.15,  0.00) {\textsc{ag\_d}\\DeepSeek};
\node[agent] (k) at (-5.15, -0.78) {\textsc{ag\_k}\\Kimi};
\node[font=\scriptsize\itshape, anchor=south] at (-5.15, 1.15)
     {Extraction role};

\node[pool] (prof) at (-3.30, 0.0) {$P$};

\node[anchor=south west, font=\scriptsize\itshape, inner sep=1pt]
      (tlabel) at (-2.42, 0.44) {one $T$ per run};
\node[tbox] (tec)   at (-1.90,  0.25) {$T_{\mathrm{EC}}$};
\node[tbox] (tgp)   at (-0.82,  0.25) {$T_{\mathrm{GP}}$};
\node[tbox] (tlww)  at (-1.90, -0.29) {$T_{\mathrm{LWW}}^{\sigma}$};
\node[tbox] (tmono) at (-0.82, -0.29) {$T_{\mathrm{MONO}}$};
\node[draw, dotted, very thin, rounded corners=2pt, gray!60!black,
      fit={(tlabel) (tec) (tmono)}, inner xsep=3pt, inner ysep=2pt]
      (tgroup) {};

\node[pool] (canon) at (0.85, 0.0) {Canonical\\$G_E$\\typed};

\node[gate] (gate) at (3.15, 0.0) {Authorize\\$\gamma(T(P),\pi)$};

\coordinate (authend) at (4.28, 0.0);

\node[outact] (aact) at (5.42,  0.36) {action $a$};
\node[outabs] (aabs) at (5.42, -0.36) {\textsc{Abstain}};

\node[draw, dashed, rounded corners=2pt, gray!65!black,
      fit={(prof) (tgroup) (canon) (gate) (authend)},
      inner xsep=6pt, inner ysep=6pt] (authbox) {};
\node[anchor=north west, font=\scriptsize, text=gray!45!black, inner sep=1pt]
      at ($(authbox.south west)+(0.0,-0.04)$)
      {institutional protocol boundary};

\draw[arr] (q.east) -- ++(0.30,0) |- (prof.north west);
\draw[arr] (d.east) -- (prof.west);
\draw[arr] (k.east) -- ++(0.30,0) |- (prof.south west);

\draw[arr] (prof.east)   -- (tgroup.west);
\draw[arr] (tgroup.east) -- (canon.west);
\draw[arr] (canon.east)  -- (gate.west);

\draw[arr] (gate.east) -- ++(0.34,0) |- (aact.west);
\draw[arr] (gate.east) -- ++(0.34,0) |- (aabs.west);

\end{tikzpicture}
\caption{One run of the protocol, read left to right. Arrows denote the flow of data within a single run, not conversational exchanges between agents: three extraction agents each contribute one typed proposal to the proposal profile $P$; the Mapping role applies exactly one transformation $T$ (one of the four alternatives shown, chosen per experimental condition) and writes the result to the canonical evidence graph $G_E$; the Controller's authorization gate reads $G_E$ and returns either an action $a$ or \textsc{Abstain}. The dashed enclosure is the institutional protocol boundary: every write inside it is checked against the role--graph authority relation and logged. The single-extractor baseline $S_1$ is not drawn; it bypasses the transformation block and forwards the designated primary proposal unchanged.}
\label{fig:mas-protocol-inline}
\end{figure}
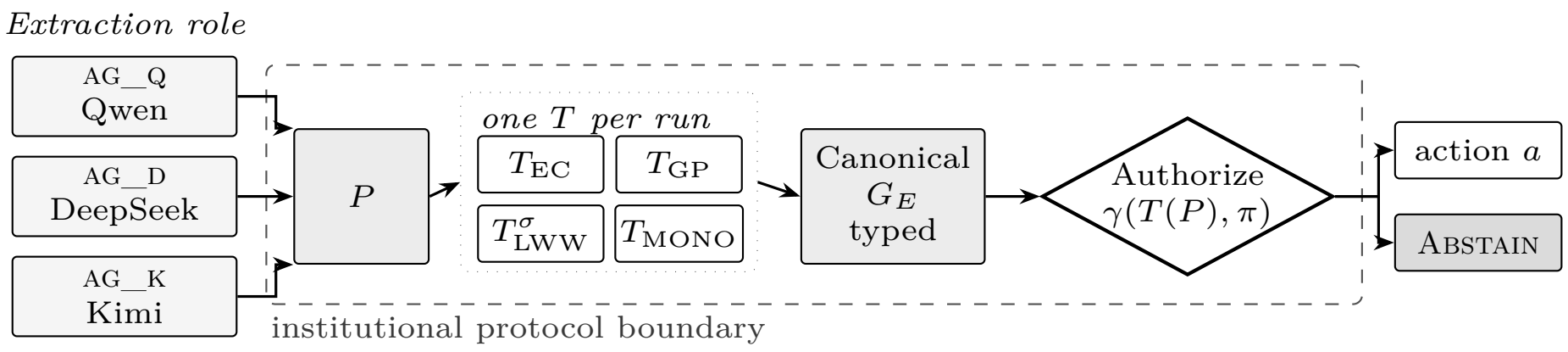

Figure~\ref{fig:mas-protocol-inline} shows the protocol and Algorithm~\ref{alg:loop} lists the procedure for a single case. We follow the running example through it, using case \texttt{CVE-2024-21762-CC01}: a synthetic benchmark variant derived from a recorded CVE advisory and constructed so that the evidence on the decision-critical field EM is in conflict (Section~\ref{sec:eval-design}).

\begin{algorithm}[t]
\caption{Governed authorization for one case}\label{alg:loop}
\begin{algorithmic}[1]
\Require evidence $e$; agents $a_1,\dots,a_k$; transformation $T$; policy $\pi$; relation $\mathsf{auth}$; trace $\tau$
\For{$i=1,\dots,k$}
  \State $s_i \gets a_i.\mathrm{propose}(e)$\label{alg:propose} \Comment{concrete, \textsc{Unknown}, or \textsc{Conflicting}}
\EndFor
\State $P \gets (s_1,\dots,s_k)$; append $P$ with provenance to $\tau$\label{alg:profile}
\State $s \gets T(P)$\label{alg:transform} \Comment{computed by the Mapping role}
\State write $s$ to $G_E$ iff $(\mathsf{role}(\textsc{Mapping}),G_E)\in\mathsf{auth}$; log it\label{alg:write}
\State $U \gets \{f : s_f \in\{\textsc{Unknown},\textsc{Conflicting}\}\}$\label{alg:unresolved} \Comment{unresolved}
\State $\mathcal{C}(s) \gets \{x\in\textstyle\prod_f V_f^* : x_f = s_f \text{ for all } f\notin U\}$ \Comment{completions}
\State $\mathsf{Act}(s) \gets \{\pi(c) : c\in\mathcal{C}(s)\}$ \Comment{action image}
\If{$|\mathsf{Act}(s)|=1$} \State \Return the unique $a\in\mathsf{Act}(s)$, logged to $\tau$
\Else \State \Return \textsc{Abstain}, logged to $\tau$\label{alg:end}
\EndIf
\end{algorithmic}
\end{algorithm}

\paragraph{Lines~1--\ref{alg:profile}: typed proposals.} The three extraction agents, backed by Qwen, DeepSeek, and Kimi, fill the same five-field schema. Each agent is called separately with an identical prompt. For EM they return \textsc{Confirmed}, \textsc{Confirmed}, and \textsc{Unknown}. Because \textsc{Unknown} is a typed value rather than a missing entry, the profile $P$ records that one agent found the evidence silent on EM.

\paragraph{Lines~\ref{alg:transform}--\ref{alg:write}: the transformation and the write.} Under field-wise majority (EC3), \textsc{Unknown} counts as an ordinary vote and loses, so the Mapping role writes EM${}={}$\textsc{Confirmed}. Under governed preservation (GP3), any non-unanimous field becomes \textsc{Conflicting}, so it writes EM${}={}$\textsc{Conflicting}. In both conditions the writer is the authorized Mapping role, so the enforcer admits the write; a write by any other role would be rejected and leave the state unchanged (Theorem~B).

\paragraph{Lines~\ref{alg:unresolved}--\ref{alg:end}: the gate.} After EC3 the completion-action image is $\{\textsc{ImmediatePatch}\}$, and the gate authorizes it, which is a false authorization on this case. After GP3, EM ranges over its full vocabulary, the image is $\{\textsc{ImmediatePatch}$, $\textsc{UrgentPatch}$, $\textsc{Monitor}\}$, and the gate abstains. Table~\ref{tab:strict-witness} in Section~\ref{sec:e3} records both traces. The same procedure, with the transformation as the only variable, generates every E3 result.

\paragraph{Vocabulary.} Table~\ref{tab:glossary} collects the condition names and metrics used throughout. The trailing digit in EC3, GP3, LWW3, and MONO3 is the number of proposal agents, not an experiment index.

\begin{table}[t]
\centering
\caption{Glossary of conditions and metrics.}
\label{tab:glossary}
\small
\begin{tabular}{@{}l>{\raggedright\arraybackslash}p{8.0cm}@{}}
\toprule
Term & Meaning \\
\midrule
S1 & single-extractor baseline: forward the primary (Qwen) proposal unchanged \\
EC3 & early collapse: field-wise typed majority vote, Qwen breaks three-way ties \\
GP3 & governed preservation: keep unanimous values, write \textsc{Conflicting} otherwise \\
LWW3 & last-writer-wins: the last arriving proposal replaces the state (six arrival orders) \\
MONO3 & order-invariant comparator: a single concrete value overrides \textsc{Unknown} \\
Completion set $\mathcal{C}(s)$ & all concrete states that agree with $s$ on its resolved fields \\
Action image $\mathsf{Act}(s)$ & set of policy actions over $\mathcal{C}(s)$; the gate acts iff it is a singleton \\
Should-abstain case & case whose oracle verdict is \textsc{Abstain} (12 of 36 in E3) \\
Oracle-decisive case & case whose oracle verdict is an action (24 of 36 in E3) \\
False authorization (FA) & a decisive verdict on a should-abstain case \\
Coverage & decisive verdicts on oracle-decisive cases, divided by 24 \\
Policy accuracy & correct actions plus correct abstentions, divided by 36 \\
\bottomrule
\end{tabular}
\end{table}

\section{Protocol and Formalization}\label{sec:protocol}
This section formalizes the three objects of Algorithm~\ref{alg:loop}: the proposal profile recording what each extraction agent typed for one case, the canonical state that a single authorized writer derives from that profile, and the gate verdict on whether that state permits an action. The institution's roles, graphs, and write authority exist so that this derivation is attributable and replayable. We present the components in dependency order: the institution and its graphs, the agents that populate them, the gate, the transformations that feed the gate, and the guarantees that follow.

\subsection{Institutional specification}
\label{sec:spec}
We model the fixed-role system as $\mathcal{M}=\langle A,\mathcal{R},G,\mathsf{auth},\mathsf{canon},\pi,\gamma,\tau\rangle$. $A$ is the agent roster, $\mathcal{R}$ the role set with assignment $\mathsf{role}:A\to\mathcal{R}$, $\mathsf{auth}\subseteq\mathcal{R}\times\{G_S,G_E,G_I,G_C\}$ the role--graph authorization relation, $\pi$ the policy, $\gamma$ the authorization gate, and $\tau$ the append-only trace; $\mathsf{canon}$ is the canonical-state transformation, denoted $T$ hereafter (Section~\ref{sec:transform}). Roles are \textsc{Retrieval}, \textsc{Extraction}, \textsc{Mapping}, \textsc{Controller}, and \textsc{Generator}. The state is partitioned into four named graphs: schema $G_S$, evidence $G_E$, inference $G_I$, and generation $G_C$. $G_S$ is read-only at runtime; \textsc{Mapping} alone writes $G_E$, \textsc{Controller} writes $G_I$, and \textsc{Generator} writes $G_C$. Every write is checked against $(\mathsf{role}(w),G')\in\mathsf{auth}$ for the requesting writer $w$ and target graph $G'$, and retained in trace $\tau$. The enforcer applies a conventional protection boundary: authenticate the writer identity and authorize the requested graph operation~\cite{saltzer1975protection}; permissions are assigned to institutional roles, following the central role-based access control (RBAC) abstraction~\cite{sandhu1996rbac}. The single-writer assignment is our protocol choice and forms the operational boundary of Theorem~B rather than a learned module. The protocol is closed and blackboard-mediated; it does not claim to solve open membership, negotiation, sanctions, or topology adaptation.

The ontology supplies finite attribute vocabularies and explicit epistemic sentinels. OWL~2 supplies the formal class and property vocabulary, while SHACL expresses constraints over the RDF graphs~\cite{w3c2012owl,w3c2017shacl}. A \textsc{Known} value has singleton completion set, so its admissible completions coincide with its declared value; \textsc{Unknown} means any concrete value in the field vocabulary remains possible; \textsc{Conflicting} retains incompatible or non-unanimous proposal provenance and treats the field as ranging over the full vocabulary. The vocabulary is enumerated, but this epistemic interpretation is not a database closed-world assumption: it uses a possible-worlds reading of incomplete information~\cite{imielinski1984incomplete,abiteboul1991possible}. A typed dictionary can implement the same semantics without a triple store; the ontology adds declarative validation, authority metadata, provenance, and queryability.

\subsection{Agents, backends, and roles}\label{sec:entities}
The protocol distinguishes three notions that are often conflated in LLM-MAS papers: an \emph{agent instance} carries identity, a local proposal buffer, and protocol credentials; an \emph{LLM backend} is the model implementation that produces natural-language or typed outputs in response to a prompt; a \emph{role} is the institutional authority bound to that instance, namely what graphs it may read or write and what messages it may emit. Roles are fixed at compile time and are not learned; backends can vary across runs without changing role assignments; instances are concrete runtime entities that own a backend.

In the E3 instantiation, three \textsc{Extraction}-role agent instances \texttt{ag\_Q}, \texttt{ag\_D}, and \texttt{ag\_K} are bound to the Qwen, DeepSeek, and Kimi backends, respectively. Each emits a typed proposal over the same evidence schema together with its provenance triple. A fourth \textsc{Mapping}-role instance \texttt{ag\_M} is the single canonical-state writer: it reads the three proposals, applies the organizational transformation $T$ (Section~\ref{sec:transform}), and writes the resulting $G_E$. The \textsc{Controller}-role instance \texttt{ag\_C} runs the authorization gate over $G_E$ and emits either an action or \textsc{Abstain}. Under the fixed E3 assignment, each instance is bound to exactly one backend, so the role--backend distinction is descriptive rather than exercised: heterogeneous disagreement equals cross-backend disagreement. Backend decoding settings and case composition are reported in Section~\ref{sec:eval-design}.

\subsection{Completion-aware authorization}
\label{sec:auth}
Let $U(s)$ be the set of fields carrying the \textsc{Unknown} or \textsc{Conflicting} sentinel in state $s$, and let $V_f^*$ denote the concrete vocabulary of field $f$. Define the admissible completion set as \[ \mathcal{C}(s)=\bigl\{x\in\prod_f V_f^*:\forall f\notin U(s),\ x_f=s_f\bigr\}, \] i.e.\ the set of full concrete states consistent with every \textsc{Known} field of $s$, with the unknown fields free to vary over $V_f^*$. The gate then computes $\mathsf{Act}(s)=\{\pi(c):c\in\mathcal{C}(s)\}$ and returns the unique element of $\mathsf{Act}(s)$ when $|\mathsf{Act}(s)|=1$ and \textsc{Abstain} otherwise. Thus uncertainty leads to abstention only when it changes the action; otherwise the gate returns a decision.

\textbf{Proposition 1 (authorization criterion).} If the gate authorizes $a$, then every $c\in \mathcal{C}(s)$ maps to $a$. \emph{Proof.} Authorization is defined by $\mathsf{Act}(s)=\{a\}$. \qed

\textbf{Corollary 1.1 (soundness).} If the true state lies in $\mathcal{C}(s)$, executing the authorized action $a$ is sound. \emph{Proof.} Proposition~1 gives $\{a\}=\{\pi(c):c\in\mathcal{C}(s)\}$, so $\pi(c)=a$ for every admissible completion. \qed

\textbf{Proposition 2 (action-level maximal permissiveness).} If all completions yield one action, authorizing it is the most permissive sound deterministic response. If two completions yield different actions, no sound deterministic gate can select one. \emph{Proof.} A selected action cannot equal two distinct policy outputs simultaneously. \qed

The guarantee assumes a fixed deterministic policy, complete admissible domains, a true state inside the completion set, correct \textsc{Known} labels, and a trusted enforcement layer. Fabricated \textsc{Known} values and a compromised authorized writer are outside the guarantee boundary and are evaluated as empirical failure modes.

\subsection{Proposal profile and organizational transformation operators}\label{sec:transform}
Let $\mathcal{A}=\{1,\ldots,k\}$ index extraction-role agent instances and let each agent emit a typed proposal $s_i\in\Sigma$ over the same evidence schema. The \emph{proposal profile} is the tuple $P=(s_1,\ldots,s_k)$. An \emph{organizational transformation} $T:\Sigma^k\to\Sigma$ collapses the profile into a single shared canonical state. The E3 instantiation evaluates a single-extractor baseline $S_1$ against four organizational transformations of the heterogeneous proposals, each defined below. With $k=3$ proposal agents these are EC3, GP3, LWW3, and MONO3, where the trailing digit counts the proposal agents rather than naming the experiment.

\paragraph{A ladder in how much disagreement survives.} The \emph{single-extractor baseline} $S_1(P)=s_1$ forwards the proposal of the designated primary extractor (Qwen); no other agent's output is consulted, so disagreement is neither collapsed nor preserved, and $S_1$ is both arrival-order invariant and identity-bound. \emph{Field-wise typed early collapse} (EC3; $T_{\mathrm{EC}}$) consults all three proposals but resolves them: for each evidence field $f$, $T_{\mathrm{EC}}(P)_f$ applies a typed majority vote over the three values, treating \textsc{Unknown} and the field-typed \textsc{Conflicting} sentinel as ordinary votes; if one value occurs at least twice that value is written, and if all three values differ the primary Qwen proposal supplies the deterministic identity-bound tie-break. EC3 is thus arrival-order invariant but not agent-symmetric, because a three-way tie depends on the designated primary identity. \emph{Governed preservation} (GP3; $T_{\mathrm{GP}}$) declines to resolve: for each field $f$ it retains the value when all three typed values are identical, the unanimous value being concrete, \textsc{Unknown}, or field-typed \textsc{Conflicting}, and rewrites any typed disagreement as the field-typed \textsc{Conflicting} sentinel, after which the completion-aware gate determines whether the disagreement crosses an action boundary.

\paragraph{Two diagnostics rather than candidate designs.} \emph{Arrival-order last-writer-wins} (LWW3; $T_{\mathrm{LWW}}^{\sigma}$) merges nothing: given a permutation $\sigma\in\mathfrak{S}_k$, $T_{\mathrm{LWW}}^{\sigma}(P)=s_{\sigma(k)}$ on the whole state, where $\sigma$ is the agent arrival order observed at the blackboard and the six replays in E3 enumerate all permutations of $\{Q,D,K\}$, the Qwen, DeepSeek, and Kimi backends. It isolates what arrival order alone can change. The \emph{order-invariant type-promoting comparator} (MONO3; $T_{\mathrm{MONO}}$) removes that dependence: field-wise, if any proposal records field-typed \textsc{Conflicting}, or at least two distinct concrete values occur, output \textsc{Conflicting}; if exactly one distinct concrete value occurs, output that value even when \textsc{Unknown} is also present; if no concrete value occurs but \textsc{Unknown} is present, output \textsc{Unknown}. The comparator is order-invariant and satisfies the stated algebraic properties, but the concrete-over-\textsc{Unknown} case can shrink the completion set, so it shows that removing order dependence does not by itself yield safety. Table~\ref{tab:operator-properties} summarizes the order, identity, and disagreement behavior of the implemented operators.

\begin{table}[t]
\centering
\caption{Implemented transformation properties. \emph{Order-inv.}: invariance to backend arrival order. \emph{Identity-bound}: a designated backend affects the result regardless of its arrival position. \emph{Diff.\ $\Rightarrow$ Conflict}: any field disagreement is mapped to the typed \textsc{Conflicting} sentinel.}
\label{tab:operator-properties}
\small
\begin{tabular}{lccc>{\raggedright\arraybackslash}p{3.6cm}}
\toprule
Operator & Order-inv. & Identity-bound & Diff.\ $\Rightarrow$ Conflict & Behavior on disagreement \\
\midrule
S1   & \checkmark & \checkmark & --        & uses one designated proposal \\
EC3  & \checkmark & \checkmark & --        & majority or primary tie-break may collapse disagreement \\
GP3  & \checkmark & --         & \checkmark & retains unanimous typed values; maps every non-unanimous field to typed conflict \\
LWW3 & --         & --         & --        & final state is the last proposal in $\sigma$ \\
MONO3 & \checkmark & --         & --        & one concrete value can dominate \textsc{Unknown} \\
\bottomrule
\end{tabular}
\end{table}

\subsection{Organizational behavior}
The realized action under transformation $T$ and fixed policy $\pi$ is
\[
\Gamma_T(P)=\gamma\bigl(T(P),\pi\bigr),
\]
where $\gamma$ returns the unique element of $\pi\!\bigl(\mathcal{C}(T(P))\bigr)$ when $|\pi(\mathcal{C}(T(P)))|=1$ and \textsc{Abstain} otherwise. E3 measures the four transformed realizations $\Gamma_{\mathrm{EC}}$, $\Gamma_{\mathrm{GP}}$, $\Gamma_{\mathrm{LWW}}^{\sigma}$ (six $\sigma$s), and $\Gamma_{\mathrm{MONO}}$ over the same replay profile and fixed policy, together with the $\Gamma_{S_1}$ baseline that bypasses the transformation block. The recorded proposal profile, schema, policy, gate, authority assignment, and case set are held fixed, so only the transformation varies.

\subsection{Protocol guarantees and organizational consequences}
\label{sec:theorems}
The principal formal claims are conditional on the preconditions listed in Section~\ref{sec:auth}; we use them to separate organizational from per-instance causes.

\paragraph{Completion refinement.}
For transformations $T_1,T_2:\Sigma^k\rightarrow\Sigma$, define
\[
T_1\preceq_C T_2
\quad\Longleftrightarrow\quad
\forall P,\;
\mathcal{C}(T_1(P))\subseteq\mathcal{C}(T_2(P)).
\]
Thus $T_1$ retains at least as much concrete information as $T_2$,
whereas $T_2$ admits at least as many possible worlds.

\textbf{Theorem A (authorization antitonicity).}
Assume nonempty completion sets and a deterministic total policy $\pi$.
If $T_1\preceq_C T_2$, then, for every profile $P$ and action $a$,
\[
\Gamma_{T_2}(P)=a
\quad\Longrightarrow\quad
\Gamma_{T_1}(P)=a .
\]

\emph{Proof.}
Let $C_i=\mathcal{C}(T_i(P))$. From $T_1\preceq_C T_2$,
$C_1\subseteq C_2$, hence $\pi(C_1)\subseteq\pi(C_2)$.
If $\Gamma_{T_2}(P)=a$, then $\pi(C_2)=\{a\}$.
Because $C_1$ is nonempty and $\pi$ is total,
$\pi(C_1)$ is a nonempty subset of $\{a\}$ and therefore equals
$\{a\}$. Thus $\Gamma_{T_1}(P)=a$. \qed

\textbf{Corollary A.1 (GP completion upper envelope).}
For every implemented transformation
\[
T\in\{S_1,T_{\mathrm{EC}},
      T_{\mathrm{LWW}}^{\sigma},T_{\mathrm{MONO}}\},
\]
we have
\[
T\preceq_C T_{\mathrm{GP}}.
\]
Consequently, every action authorized after GP3 is authorized as
the same action after each of these transformations; the converse
need not hold.

\emph{Proof.}
Consider a field $f$. If all proposals agree, GP3 and every
transformation above retain that unanimous typed value. Otherwise,
GP3 writes $\textsc{Conflicting}_f$, whose completion set is the full
concrete domain $V_f^*$. Any concrete value or epistemic sentinel
written by $S_1$, EC3, LWW3, or MONO3 admits a subset of $V_f^*$.
Field-wise containment and Cartesian composition therefore give
\[
\mathcal{C}(T(P))\subseteq
\mathcal{C}(T_{\mathrm{GP}}(P)).
\]
The authorization result follows from Theorem~A. \qed

Being the upper envelope makes $T_{\mathrm{GP}}$ the least permissive implemented operator. Because its authorization set is contained in that of every other operator, its lowest decisive coverage is structural. Its minimal false-authorization count is guaranteed by construction when the oracle's should-abstain labeling matches the policy's decision boundary; otherwise it is an empirical outcome on this benchmark.

Strictness at profile $P$ occurs exactly when
$|\pi(\mathcal{C}(T(P)))|=1$ but
$|\pi(\mathcal{C}(T_{\mathrm{GP}}(P)))|>1$.

\textbf{Theorem B (rejected-write invariance).} If a writer $w$ on graph $G'$ satisfies $\neg((\mathsf{role}(w),G')\in\mathsf{auth})$ and the enforcer rejects the transition, then for any transformation $T$ and profile $P$ the resulting canonical state, completion set, and authorization are invariant: $\Gamma_T(P)$ depends only on canonical-state changes admitted by $\mathsf{auth}$. This is a local access-control integrity property for rejected transitions; it does not claim information-flow noninterference.

\emph{Proof sketch.} The rejected transition has no effect on any canonical state; the transformation $T$ depends only on its arguments; the gate depends only on the post-canonical state; therefore the entire realization chain is invariant under the rejected write. \qed

\textbf{Theorem C (order invariance).} For each field, the binary operator induced by MONO3 is the join on a finite carrier with \textsc{Unknown} as bottom, concrete values as incomparable atoms, and \textsc{Conflicting} as top. It is therefore associative, commutative, and idempotent (ACI), and the MONO3 ACI algebra makes $\Gamma_{\mathrm{MONO}}(P)$ invariant under agent arrival order. This algebraic order is not completion-set inclusion: joining one concrete with \textsc{Unknown} returns the concrete. By contrast, $T_{\mathrm{LWW}}^{\sigma}$ realizes $x\diamond y=y$, which is associative and idempotent but not commutative; different arrival orders can therefore select different canonical states. E3 evaluates all six permutations of $\{Q,D,K\}$, while the implementation tests exhaust the finite MONO3 carriers. Each formal result links a design choice to an experimental observable: Theorem~A and Corollary~A.1 to the GP3--EC3 coverage delta, Theorem~B to the authority intervention, and Theorem~C to the LWW3 permutation span.

\section{Related Work}\label{sec:related}
Adjacent work spans heterogeneous LLM proposals, normative MAS organizations, incomplete-information combination, selective prediction, runtime assurance, and ontology governance.

\paragraph{Heterogeneous proposal and aggregation.} Surveys of LLM-based MAS organize the field around agent profiling, communication, and capability acquisition~\cite{guo2024llmmultiagents}. Multiagent debate exchanges responses to improve reasoning and factuality~\cite{du2024multiagentdebate}; formal analysis also identifies majority-convergence and common-misconception failure modes~\cite{estornell2024multillm}. Separately, black-box LLM evaluations measure whether a model withholds an answer when uncertain or when no supported answer is available~\cite{madhusudhan2025abstention}, while uncertainty-aware language agents use quantified uncertainty to mediate interaction with external tools or environments~\cite{han2024uala}. Ontology-grounded knowledge-graph construction aligns extracted relations to an external schema~\cite{feng2024ontologygrounded}; the Extract--Define--Canonicalize (EDC) framework instead performs open extraction, schema definition, and post-hoc canonicalization~\cite{zhang2024edc}. These methods operate at the response, tool-interaction, or schema-construction layer. The present protocol differs in three respects: the proposal profile carries epistemic type; the transformation $T$ is explicit; and the action boundary is the gate's verdict rather than an LLM selection.

\paragraph{Electronic institutions and organizational models.} Electronic-institution languages formalize norms and obligations~\cite{esteva2001electronic}, organizational models derive interaction from role-based logical specifications~\cite{dignum2004model}, and coordination artifacts mediate heterogeneous agents through their environment~\cite{omicini2004artifacts}. Our named-graph blackboard instantiates these abstractions. Whereas recent agent-security work enforces privilege policies on tool calls~\cite{progent2025}, our enforcer mediates canonical-graph writes and Theorem~B states the rejected-write invariant.

\paragraph{Incomplete information and belief combination.} Epistemic accounts distinguish individual from common knowledge~\cite{halpern1990knowledge}; incomplete databases represent partial information by sets of possible worlds~\cite{imielinski1984incomplete,abiteboul1991possible}; belief merging combines bases under integrity constraints~\cite{konieczny2002merging}. We operationalize only possible-world completion: the gate enumerates policy outcomes over $\mathcal{C}(s)$, without proposing a new epistemic logic or general merging operator.

\paragraph{Selective prediction, runtime assurance, and ontology governance.} Selective classification formalizes risk--coverage~\cite{elyaniv2010selective}, and \textsc{SelectiveNet} learns it end to end~\cite{geifman2019selectivenet}. LLM abstention spans query answerability, model confidence, and value alignment~\cite{wen2025abstention}; our gate addresses the first through explicit completions. Conformal prediction provides distribution-free prediction sets under exchangeability~\cite{angelopoulos2023conformal}, statistically complementing our deterministic gate. Runtime verification monitors finite traces~\cite{barringer2004rule}, including inconclusive three-valued verdicts~\cite{bauer2011runtime}. Ontologies can augment partial observations~\cite{ghanadbashi2023observation} or inform prediction and explanation~\cite{smirnov2024ontology}; here the ontology serves as a finite-vocabulary governance contract.

\paragraph{Positioning.} The nearest neighbours each cover one step of Algorithm~\ref{alg:loop}. Voting and debate implement a particular choice of $T$ (EC3 is one such choice) but treat the merged output as final. Institutional and access-control models implement line~\ref{alg:write} but say nothing about the content of an authorized write. Selective prediction, conformal sets, and three-valued runtime monitors implement a form of lines~\ref{alg:unresolved}--\ref{alg:end}, but over a model's confidence, a statistical prediction set, or a temporal property, rather than over the completions of a shared state under an explicit policy. Belief merging studies $T$ in general but not its downstream authorization consequences. These lines of work could be extended or combined to reproduce parts of the protocol; our claim is narrower. By fixing a common semantics, namely typed fields, completion sets, and a deterministic policy, the protocol makes merge rules comparable by a single order ($\preceq_C$) whose authorization consequence is provable (Theorem~A) and checkable on each recorded case (Section~\ref{sec:e3}). In the running example, an event trace of the system would show two well-formed, authorized writes; only the completion-action image explains why one verdict is unsafe.

\section{Evaluation Design}
\label{sec:eval-design}
All experiments use recorded inputs and deterministic post-processing; no new LLM calls are made for E3. Statistical units are case clusters rather than call counts. The protein-policy benchmark used in E1--E2 is set in promoter selection for protein engineering, where a construct is chosen by weighing factors on which the evidence may be silent or inconsistent: it is a synthetic case set over a controlled protein-domain ontology whose five attributes are expression demand, toxicity, folding burden, localization, and cofactor status. Each case pairs an evidence record with a written policy that names the admissible action per attribute state, and the four E2 missingness patterns are layered on the same ontology, so a missing or contested attribute stands for evidence that is silent or inconsistent on that factor. E1 uses 24 cases. E2 uses a 48-case authorization benchmark and a separate 36-case clarity benchmark: the former scores the authorization decision, the latter scores state integrity before authorization. E3 uses a fixed 36-case CVE replay set, of which 12 should abstain and 24 are oracle-decisive; the 12 should-abstain cases consist of six conflicting-critical and six unsupported-critical variants, and the 24 oracle-decisive cases consist of 12 \textsc{Known} cases, six conflicting-dominated variants, and six unsupported-dominated variants. Each synthetic variant identifier appends a variant suffix to the base Common Vulnerabilities and Exposures (CVE) record: \texttt{C} for unsupported-critical, \texttt{CC} for conflicting-critical, \texttt{D} for unsupported-dominated, and \texttt{CD} for conflicting-dominated; any trailing digits index individual variants within a variant class (Table~\ref{tab:strict-witness}). Coverage is the number of oracle-decisive cases answered decisively, divided by 24; policy accuracy is the number of correct decisive actions plus correct abstentions, divided by 36. False authorization is decisive output on a should-abstain case. E3 holds the proposal profile, schema, policy, gate, authority topology, and case set fixed while varying only the organizational transformation. The three E3 backends are Qwen \texttt{qwen3.7-plus} (temperature $0.1$, top-$p$ $1.0$), DeepSeek \texttt{deepseek-v4-flash} (temperature $0.1$, top-$p$ $1.0$), and Kimi \texttt{kimi-k2.6} (temperature $0.6$, top-$p$ $0.95$, reasoning disabled); all three receive an identical prompt over the same five-field schema.

\subsection{Replay and unit semantics}\label{sec:replay}
Each E3 case is an atomic cluster: the same three backend proposals are replayed deterministically across the single-extractor baseline, the four transformations, and the six LWW arrival orders. The total replay count is $36 \times (1 + 1 + 1 + 6 + 1) = 360$ organizational realizations, none of which re-invokes an LLM. LLM call counts and LWW permutations are not independent statistical units; headline outcomes are case-cluster counts, whereas status and disagreement survival use explicitly reported field-level eligible-record denominators. E1 aggregates across repeats as stated in Section~\ref{sec:e1}, and E2 reports balanced accuracy on the 48-case benchmark without folds or shuffles.

\subsection{Mechanism reconstruction}\label{sec:provenance}
For profile $P$, the raw disagreement fields are $\Delta(P)=\{f:|\{s_i(f)\}_i|>1\}$. Status survival counts field records containing an input sentinel whose transformed value remains a sentinel; disagreement survival counts fields in $\Delta(P)$ that the transformation maps specifically to \textsc{Conflicting}. Together they reconstruct the chain from proposal disagreement through canonical transformation to gate verdict without an oracle target field or a new model call. Their denominators are field records, not cases.

\section{E1: Coordination and Policy Dependence}
\label{sec:e1}
E1 combines the controlled-vocabulary coordination benchmark and the policy ablation. The E1 evidence chain uses a single LLM backend (Kimi), 24 cases, three vocabulary conditions with three repeats each, and five policy conditions with five repeats on the LLM paths. Across this trace, the full vocabulary reaches 0.989 mean state agreement with zero schema violations by construction and 0.958 inter-run consistency across the three repeats; partial vocabulary reaches 0.925 agreement with 0.400 schema violations and 1.000 consistency; free-form coordination reaches 0.772 agreement with 0.961 violations and 0.833 consistency. State agreement and schema violation are means over the 24 cases $\times$ 3 repeats of the per-run fraction of the five attributes; inter-run consistency is the fraction of the 24 cases whose three repeats yield an identical canonical state. We retain the artifact condition identifiers \texttt{C1}, \texttt{C1-JSON}, \texttt{C2}, \texttt{C3}, \texttt{C3-P} for the deterministic explicit-policy path, its dictionary-state equivalent, the implicit-policy LLM path, the no-policy LLM path, and the explicit-policy LLM path, respectively. The closed vocabulary that the ontology declares reduces semantic drift at the representation boundary; whether that boundary translates into downstream decisions is the question the policy ablation answers. Multi-backend generalization is not claimed at the E1 layer; heterogeneous-backend disagreement is the E3 observation reported in Section~\ref{sec:e3}.

The deterministic rule path is 100\% exact on 24 cases; C1-JSON is the same rule engine over a plain-dict state and is exact by the same construction, so the RDF carrier is not a decision determinant. On the LLM paths, the denominator is 120 runs (24 cases $\times$ 5 repeats): the LLM reading raw evidence without the policy recovers only 25.8\% of oracle actions (C2 implicit) or 24.2\% from the structured state (C3 no-policy) and over-strengthens 55.0\% or 30.8\% respectively; providing the complete policy raises exact recovery to 100\% (C3-P) with 0.0\% over-strength. Explicit policy execution is the operative determinant. The contrast between policy-bound and policy-absent paths motivates E2's epistemic-authorization study. Table~\ref{tab:e1-compact} summarizes both vocabulary and policy panels.

\begin{table}[t]
\centering
\caption{E1 coordination and policy dependence, single backend (Kimi). \textbf{(a)} Vocabulary conditions, 24 cases $\times$ 3 repeats. \textbf{(b)} Policy conditions, 24 cases (deterministic paths) and 120 runs (LLM paths).}
\label{tab:e1-compact}
\small
\begin{tabular}{lrrr}
\toprule
\textbf{(a)} Vocabulary condition & State agreement & Schema violation & Inter-run consist. \\
\midrule
Ontology full vocabulary & 0.989 & 0.000 & 0.958 \\
Partial vocabulary & 0.925 & 0.400 & 1.000 \\
Free-form & 0.772 & 0.961 & 0.833 \\
\bottomrule
\end{tabular}

\vspace{6pt}

\begin{tabular}{lrr}
\toprule
\textbf{(b)} Policy condition & Exact recovery & Over-strengthening \\
\midrule
C1 deterministic explicit & 1.000 & 0.000 \\
C1-JSON dictionary & 1.000 & 0.000 \\
C2 LLM implicit & 0.258 & 0.550 \\
C3 LLM no policy & 0.242 & 0.308 \\
C3-P LLM explicit policy & 1.000 & 0.000 \\
\bottomrule
\end{tabular}
\end{table}

\section{E2: Epistemic Authorization}
\label{sec:e2}
E2 evaluates 48 cases crossing irrelevant, dominated, critical, and conflicting missingness, of which 24 are answerable and 24 are unanswerable. The closed-world and gate-less baselines each obtain balanced accuracy 0.500 (the mean of recall on answerable and unanswerable cases) because they answer 24/24 unanswerable cases rather than abstaining. We retain the artifact condition identifiers \texttt{V1}, \texttt{V2-U}, \texttt{V2-A}, \texttt{D2}, \texttt{O2} for closed-world coercion, sentinels without the gate, sentinels with the gate, a Python-domain dictionary, and OWL-declared domains, respectively. \texttt{D2} and \texttt{O2} resolve to the same loaded vocabulary on the present ontology, so their agreement is an invariance property of the gate rather than an independent replication; they are retained to show the gate is robust to that surface change. Epistemic status becomes operational through the completion-aware gate; without the gate it is an inert label. The accuracy gain comes from the gate itself; RDF serves as the institutional carrier.

\begin{table}[t]
\centering
\caption{E2 authorization and state integrity. \textbf{(a)} Balanced accuracy by authorization condition on the 48-case authorization benchmark (24 answerable, 24 unanswerable); V1 and V2-U answer every case and obtain the degenerate always-answering baseline. D2 and O2 differ only in where the value domains are declared and resolve to the same loaded vocabulary, so their equality is an invariance property of the gate rather than an independent replication. \textbf{(b)} State-integrity aggregate on the separate 36-case clarity benchmark (24 answerable, 12 unanswerable).}
\label{tab:e2-compact}
\small
\begin{tabular}{lr}
\toprule
\textbf{(a)} Authorization condition & Balanced accuracy \\
\midrule
V1 closed-world coercion, always answers & 0.500 \\
V2-U explicit sentinels, no gate & 0.500 \\
V2-A explicit sentinels with gate & 1.000 \\
D2 gate, dictionary-sourced domains & 1.000 \\
O2 gate, ontology-sourced domains & 1.000 \\
\bottomrule
\end{tabular}

\vspace{6pt}

\begin{tabular}{lr}
\toprule
\textbf{(b)} State-integrity metric & Value \\
\midrule
Balanced accuracy (overall) & 0.896 \\
Conflict detection & 0.917 \\
Unknown recognition & 1.000 \\
Value accuracy on known fields & 0.833 \\
Unsafe answers (count) & 0 \\
\bottomrule
\end{tabular}
\end{table}

The 0.500 balanced accuracy of V1 and V2-U is the degenerate always-answering baseline; the 1.000 balanced accuracy of V2-A, D2 and O2 is the gate at work. The 0.896 state-integrity balanced accuracy (Table~\ref{tab:e2-compact}) is computed on the separate 36-case clarity benchmark and measures extraction quality before authorization: \textsc{Unknown} recognition is 1.000, conflict detection is 0.917, and value accuracy on \textsc{Known} fields is 0.833. Authority enforcement is evaluated separately below.

The complementary authority intervention exercises the ontology-declared role--graph relation, which the enforcer loads from the schema at runtime, by injecting cross-role writes over 24 evidence records and 6 tamper records; it blocks 72/72 injected writes under partitioned enforcement versus 0/72 without enforcement, and also verifies 24/24 traces and detects 6/6 tamper cases. These figures characterize the authority layer; they are not a CVE performance result.

\section{E3: Organizational Preservation}
\label{sec:e3}
E3 instantiates the operators of Section~\ref{sec:transform} on the fixed replay profile of Section~\ref{sec:eval-design}: S1 forwards the designated primary proposal unchanged; EC3 is field-wise typed majority with Qwen as deterministic identity-bound tie-break; GP3 maps any field disagreement to that field's typed \textsc{Conflicting} sentinel; LWW3 applies whole-state last-writer-wins under all six arrival orders; MONO3 is an order-invariant associative, commutative, and idempotent (ACI) diagnostic.

\begin{table}[t]
\centering
\caption{Organizational preservation on 36 recorded proposal profiles. Metric definitions are given in Section~\ref{sec:replay}. FA, abst., acc. denote false authorization, abstention, and accuracy, respectively. $^{\dagger}$LWW3 reports the spread over six arrival-order replays, which are repeated views of the same cases rather than independent observations; the full min--max range over replays is 2--4/12. $^{\ddagger}$MONO3 is a mechanism diagnostic.}
\label{tab:mas-organization}
\small
\begin{tabular}{lccccc}
\toprule
Condition & FA (/12) & Correct abst. (/12) & Coverage & Policy acc. & Decisive (/36) \\
\midrule
S1 & 3/12 & 9/12 & 0.792 & 0.750 & 22 \\
EC3 & 1/12 & 11/12 & 0.833 & 0.833 & 21 \\
GP3 & 0/12 & 12/12 & 0.708 & 0.806 & 17 \\
LWW3$^{\dagger}$ & 2--4/12 & 8--10/12 & 0.792--0.917 & 0.750--0.806 & 21--26 \\
MONO3$^{\ddagger}$ & 6/12 & 6/12 & 0.875 & 0.722 & 27 \\
\bottomrule
\end{tabular}
\end{table}

\begin{table}[t]
\centering
\caption{LWW3 arrival-order diagnostics. The six replays enumerate all permutations of $\{Q,D,K\}$ and are repeated views of the same 36 cases, not independent observations; the rows below report the min--max spread of the corresponding LWW3 metrics. avg/max denote the average and maximum diversity across the six replays.}
\label{tab:mas-lww-diagnostics}
\small
\begin{tabular}{lc}
\toprule
Diagnostic & Value (denominator) \\
\midrule
Any false-auth cases & 7/12 \\
Order-sensitive unsafe & 7/12 \\
Unsafe under all six arrival orders & 0/12 \\
Canonical-state order-sensitive & 18/36 \\
Decision-verdict order-sensitive & 11/36 \\
Canonical-state diversity (avg/max) & 1.611 / 3 \\
Decision-verdict diversity (avg/max) & 1.333 / 3 \\
\bottomrule
\end{tabular}
\end{table}

\begin{table}[t]
\centering
\caption{Mechanism-boundary transition on the same 36 profiles. Raw disagreement: 18/36 case clusters and 27 field records. Decision-relevant profile disagreement: 13/36 cases. Column denominators are stated in the header; metric definitions are in Section~\ref{sec:provenance}.}
\label{tab:mas-mechanism}
\small
\begin{tabular}{lcccccc}
\toprule
Transformation & \shortstack[l]{Status kept\\(/46)} & \shortstack[l]{Mapped\\to Conflict\\(/27)} & \shortstack[l]{Multi-action\\(/36)} & \shortstack[l]{Correct\\abst. (/12)} & \shortstack[l]{False\\auth. (/12)} & \shortstack[l]{Coverage\\(/24)} \\
\midrule
S1 & 33/46 & 0/27 & 14/36 & 9/12 & 3/12 & 0.792 \\
EC3 & 33/46 & 0/27 & 15/36 & 11/12 & 1/12 & 0.833 \\
GP3 & 46/46 & 27/27 & 19/36 & 12/12 & 0/12 & 0.708 \\
MONO3 & 23/46 & 4/27 & 9/36 & 6/12 & 6/12 & 0.875 \\
\bottomrule
\end{tabular}
\end{table}

Table~\ref{tab:mas-organization} reports false-authorization, abstention, coverage, and policy accuracy for the S1 baseline and the four organizational transformations on the fixed 36-case profile. Across LWW3's six non-independent arrival-order views (Table~\ref{tab:mas-lww-diagnostics}), false authorization spans 2--4/12 and policy accuracy spans 0.750--0.806; the order-sensitive false-authorization count is 7 across the six views.

Proposal diversity becomes useful for disagreement detection only when the canonical transformation preserves decision-relevant uncertainty. E3 demonstrates this mechanism on one heterogeneous recorded profile; the effect cannot be attributed uniquely to backend heterogeneity from this single profile. Status-survival and disagreement-survival are distinct field-level diagnostics; neither is a universal preservation theorem.

\begin{table}[t]
\centering
\caption{Strict authorization witness on the synthetic conflicting-critical benchmark variant \texttt{CVE-2024-21762-CC01}; the policy and gate are identical, only the organizational transformation changes, and EM denotes exploit maturity.}
\label{tab:strict-witness}
\small
\begin{tabular}{lll}
\toprule
Stage & EC3 & GP3 \\
\midrule
Agent EM proposals
  & \multicolumn{2}{l}{Confirmed, Confirmed, Unknown} \\
Canonical EM
  & Confirmed & Conflicting \\
Completion-action image
  & $\{\textsc{ImmediatePatch}\}$
  & \begin{tabular}[t]{@{}l@{}}$\{\textsc{ImmediatePatch},$\\
      $\textsc{UrgentPatch},\textsc{Monitor}\}$\end{tabular} \\
Gate verdict
  & Authorize (false) & \textsc{Abstain} \\
\bottomrule
\end{tabular}
\end{table}

This witness realizes the strict case of Theorem~A: EC3 removes an
agent disagreement before authorization, whereas GP3 retains it in the
canonical state, enlarges the completion-action image, and changes the
collective verdict from false authorization to abstention. The policy
and gate are unchanged, so the difference is attributable to the
organizational transformation of the shared state. This case is EC3's only false authorization on the benchmark: the
single-case witness and the aggregate EC3--GP3 false-authorization
gap of one should-abstain cluster are the same event viewed at two
granularities.

\subsection{Paired analysis}\label{sec:paired}
The EC3--GP3 differences are small in both strata, so we report case-level discordant counts rather than marginal rates, and we separate the safety stratum from the coverage stratum. On the 12 should-abstain cases, EC3 false-authorizes where GP3 abstains in $1$ case and the reverse occurs in $0$ (paired McNemar, exact two-sided $p=1.0$). On the 24 oracle-decisive cases, EC3 is decisive where GP3 abstains in $3$ cases with $0$ in the reverse direction (exact two-sided $p\approx 0.25$); this is a coverage difference, not a safety difference. The count of 7 reported above in Table~\ref{tab:mas-lww-diagnostics} records LWW3 order-sensitive unsafe cases under a different transformation comparison on the same 36-case profile; it is not the EC3--GP3 paired-discordance count. Coverage Wilson intervals at the 95\% level are approximately $0.83\ (0.64,\ 0.93)$ for EC3 and $0.71\ (0.51,\ 0.85)$ for GP3. With $n=24$ clusters and a deterministic replay rather than independent sampling, these intervals bound precision and do not support a population-level superiority claim; the mechanism results below do not rest on one. The MONO3 invariant and 6/12 false-authorization result is a mechanism diagnostic: order invariance alone is not sufficient for safety.

\section{Discussion}\label{sec:discussion}
\subsection{Redundancy, transformation, and policy}
Taken together the three evaluations answer when redundancy becomes governable. E1 places decision conformance on the side of explicit policy rather than ontology, E2 disentangles epistemic gating from authority enforcement, and within the fixed replay E3 attributes the observed safety--coverage differences to the organizational transformation rather than to changes in the underlying proposals. This does not make the ontology inert: the closed vocabulary of E1 and the role--graph relation that blocks 72/72 cross-role writes in E2 are both declared in the schema and enforced at runtime. What the ontology does not supply is the decision rule. Redundancy exposes disagreement only when agents produce distinct proposals, and can reduce coverage when explicit conflict propagation makes action ambiguity visible. The relevant design choice is therefore the transformation together with the authorization policy.

\subsection{Safety--coverage trade-off and field-level diagnostics}
GP3 abstains on the safety stratum while EC3 false-authorizes one case on the same stratum; EC3 is decisive where GP3 abstains in 3 cases on the coverage stratum. LWW3 shows that unconstrained arrival order can change safety outcomes. Theorem~A and Corollary~A.1 formally relate completion sets and authorization antitonicity; the empirical paired-discordance split and the strict witness are the case-level evidence for the strict reading.

S1 retains sentinel status but never maps disagreement to \textsc{Conflicting}, and MONO3 retains partial disagreement. Table~\ref{tab:mas-mechanism} connects field transitions to completion-action images and authorization outcomes.

\subsection{Lessons for engineering LLM-MAS}\label{sec:lessons}
The following practices follow from the evidence above. Each names the result it rests on; none has been validated beyond the present system.
\begin{enumerate}[leftmargin=*,itemsep=1pt,topsep=2pt]
\item \emph{Name the merge step and review it like a component.} With proposals, policy, and gate fixed, the choice of $T$ alone moved false authorization between 0/12 and 6/12 (Table~\ref{tab:mas-organization}). A merge rule that is implicit in glue code is an unreviewed safety decision.
\item \emph{Keep \textsc{Unknown} and \textsc{Conflicting} distinct from concrete values and from each other, and do not let them be outvoted.} The only EC3 false authorization arose because an \textsc{Unknown} proposal was counted as a losing vote (Table~\ref{tab:strict-witness}).
\item \emph{Do not take order invariance as a safety argument.} MONO3 is provably order-invariant (Theorem~C) and had the most false authorizations (6/12), because a single concrete value overrode \textsc{Unknown}. Conversely, last-writer-wins made safety depend on arrival order (2--4/12).
\item \emph{Execute the decision policy over the typed state; do not ask a model to apply it from prose.} The same LLM recovered 24.2\% of oracle actions without the policy and 100\% with the executable policy (Table~\ref{tab:e1-compact}); typed sentinels without the gate were no better than always answering (Table~\ref{tab:e2-compact}).
\item \emph{Enforce write authority at the shared state, not at the agents.} Partitioned enforcement blocked 72/72 injected cross-role writes against 0/72 without it, and made the rejected-write guarantee (Theorem~B) apply.
\item \emph{Budget for the coverage cost and say who handles abstentions.} GP3's safety came with coverage 0.708 against 0.833 for EC3; in a deployment each abstention must be routed to further evidence collection or a human.
\item \emph{Add an independent check for shared errors.} Disagreement preservation is silent when every agent makes the same mistake; secondary evidence validation is still needed.
\end{enumerate}
To adopt the protocol, a system needs at least: a finite vocabulary per decision-relevant field, extended with the two sentinels; a deterministic policy over complete states; a single writer role for the shared state; and a log of proposals and writes. The RDF substrate is optional (Section~\ref{sec:spec}).

\subsection{Threats to Validity}\label{sec:threats}
The formal guarantees assume closed nonempty completion domains, a
deterministic total policy, and a trusted authority enforcer. The
empirical benchmarks measure conformance to that policy rather than
real-world CVE or biological correctness. E3 is a case-clustered
deterministic replay of one recorded 36-case proposal profile: it
isolates organizational transformations but does not estimate
population-level effects, backend independence, or superiority over
same-backend self-consistency; backend identity is also confounded with
decoding configuration. The EC3--GP3 safety difference rests on a single
should-abstain case (Section~\ref{sec:paired}); it is a mechanism witness,
not a statistically supported effect. The should-abstain designation is derived from
the same fixed policy that the gate executes, so the oracle and the
mechanism share a specification; this bounds the trade-off result to the
present benchmark. E1 uses a single backend, and the E1--E2 benchmarks are
synthetic. The system is one closed, fixed-role MAS; open membership,
negotiation, and larger or more heterogeneous agent populations are not
evaluated. Finally, disagreement preservation cannot detect a
concrete error shared by every proposal agent, for which secondary
evidence validation remains necessary.

\subsection{Future Work}\label{sec:future}
Each limitation suggests a test. First, an oracle labelled independently of the executed policy, for example by domain experts on real advisories, would show whether the safety--coverage ordering holds when oracle and gate can disagree. Second, comparing heterogeneous backends with same-backend self-consistency at matched decoding settings would separate the value of model diversity from that of mere resampling. Third, larger and adversarially drawn proposal profiles, and MAS with more roles and open membership, would test whether the lessons above transfer. Fourth, the deterministic gate could be combined with statistical prediction sets~\cite{angelopoulos2023conformal} so that completions are weighted rather than enumerated, trading part of the guarantee for coverage. Finally, routing abstentions to targeted evidence retrieval would measure how much coverage can be recovered without giving up the guarantee.

\section{Conclusion}
Useful proposals do not necessarily produce a state that is ready for action. The problem arises whenever heterogeneous agents interpret incomplete or conflicting evidence, whether in vulnerability triage or in design choices such as promoter selection for protein engineering, the setting of our E1--E2 benchmarks.

The protocol separates proposal generation, canonical transformation under role-based writing authority, and policy-bound authorization, with the ontology providing the shared vocabulary and epistemic types. Three guarantees follow from that separation: authorization is antitone in the completion set (Theorem~A), so a transformation preserving more disagreement is never more permissive; a rejected write leaves the whole realization unchanged (Theorem~B); and order invariance is attainable without being sufficient for safety (Theorem~C).

The measurements separate what the ontology fixes from what the policy decides. The schema-declared vocabulary holds state agreement at 0.989 against 0.772 for free-form coordination, and the schema-declared authority relation blocks 72/72 cross-role writes; what raises exact recovery from 24.2\% to 100\% is supplying the policy, not changing the carrier. Across the fixed CVE replay the transformation, not the number of agents, sets the safety--coverage position: GP3 maps every observed field disagreement to typed conflict and reaches 0/12 false authorization at 0.708 coverage, EC3 accepts one false authorization for three further decisive answers, and arrival order alone moves false authorization between 2/12 and 4/12.

The evidence is bounded to one recorded profile and one policy specification, and each bound maps to a direction: an oracle specified independently of the executed policy, evidence channels outside the agent pool, and proposal profiles drawn from adversarial rather than fixed agent populations. The design rule the present evidence supports is narrower and firmer: decision-relevant uncertainty must survive until a policy-bound gate can justify acting on it. An ontology-governed protocol is what makes that survival declarable, enforceable, and auditable.

\begin{credits}
\subsubsection{\ackname} Generative AI tools assisted manuscript preparation, code drafting, and statistical design. The authors verified all methods, code, results, and text and remain fully accountable for the work.
\end{credits}

\bibliographystyle{splncs04}
\bibliography{references}
\end{document}